\documentclass[a4paper,10pt]{article}
\usepackage{latexsym, amsmath, amsfonts, amssymb, array, dcolumn, hhline, longtable, enumerate,psfrag,bm}
\usepackage[english]{babel}
\usepackage[dvips]{graphicx}
\usepackage{epsfig}
\usepackage[small,bf]{caption}
\begin{document}
%
%
%

\noindent
{\bf \Large{Defect-induced ferromagnetism in fullerenes}}

\vspace{0.5 cm}

\noindent
{\bf \large {D.W. Boukhvalov and M. I. Katsnelson}}

\vspace{0.5 cm}

\noindent
{\it Institute for Molecules and Materials,
Radboud University Nijmegen,
Toernooiveld 1,
6525 ED Nijmegen, ~~~~~~~~~ ~~~~~~~~~ ~~~~~The Netherlands}

\begin{abstract}
Based on the {\it ab initio} electronic structure calculations the
picture of ferromagnetism in polimerized C$_{60}$ is proposed
which seems to explain the whole set of controversial experimental
data. We have demonstrated that, in contrast with cubic
fullerene, in rhombohedral C$_{60}$ the segregation of iron atoms
is energetically unfavorable which is a strong argument in favor
of intrinsic character of carbon ferromagnetism which can be
caused by vacancies with unpaired magnetic electrons. It is shown
that: (i) energy formation of the vacancies in the rhombohedral
phase of C$_{60}$ is essentially smaller than in the cubic phase,
(ii) there is a strong ferromagnetic exchange interactions between
carbon cages containing the vacancies,
(iii) presence of iron impurities can diminish essentially
the formation energy of intrinsic defects, and (iv) the fusion of the
magnetic single vacancies into nonmagnetic bivacancies is energetically
favorable. The latter can explain a fragility of the ferromagnetism.
\end{abstract}

%
%
 
%
\section{Introduction}
\label{intro}

Carbon plays unique role in nature, from nucleosynthesis processes
in stars to its crucial importance as a basic material substrate
of life. It is not surprising therefore that all its properties is
a subject of great interest. Magnetic properties of elemental
carbon were studied in a context of geology and cosmology
\cite{1}, biochemistry \cite{2}, physics \cite{3}, and material
sciences \cite{4}. The possibility of intrinsic long-range
magnetic order in carbon is intriguing from both basic and applied
scientific point of view since the properties of $sp$-electron
magnetic materials can be essentially different from conventional
itinerant-electron ferromagnets, in particular, it follows from
general theoretical considerations that they may be perspective
candidates to magnetic semiconductors with the Curie points $T_C$
above room temperature \cite{EK}.

The first observation \cite{mak1} the ferromagnetism with $T_C$
about 500K in polymerized fullerenes attracts much of attention and
now this is the most studied magnetic carbon-based system. The
experimental data are still very controversial and there is still
no common opinion about intrinsic or extrinsic character of this
ferromagnetism. The reported results were independently reproduced
afterwards \cite{wood,narozhn} and conditions of appearance of
the ferromagnetism in polymerized fullerenes were studied in more
detail \cite{wood,struct,photo,conduct,condit,maknew}. It was found that
the ferromagnetism arises only in a very narrow range of pressure
and temperature in the vicinity of the point where fullerene cages
are collapsed \cite{wood,condit} (for a general review of current
status of the problem, see Ref. \cite{rev}). A typical criticism
\cite{rev} which always occurs when discussing ferromagnetism in
$sp$-electron systems with small magnetization (cf. the discussion
for CaB$_6$ \cite{EK}) is a possible contamination of samples by
iron and iron-based compounds due to use of metallic tools,
however, a very careful experimental procedure avoiding this
problem is described in Ref. \cite{condit}. The most serious
criticism that have even resulted in the retraction of the first
report \cite{retract} was presented in Ref. \cite{Fe3C} where on
the base of closeness of the Curie temperatures of ferromagnetic
C$_{60}$ and cementite (Fe$_3$C) the effect was attributed to the
presence of the latter phase.
%
%
An idea of crucial role
of Fe impurities and formation of ferromagnetic Fe$_3$C phase can
not explain the disappearance of ferromagnetism beyond a very
narrow pressure and temperature interval of sample preparation
\cite{wood,narozhn,condit} since under high temperatures and pressures the
cementite decays into kish graphite and pure iron, the latter
having even higher Curie temperature than cementite. Also, it is
very difficult to explain in these terms a strong dependence of
the magnetization of photopolymerized fullerenes on the choice of
oxygen-free or oxygen-rich formation conditions \cite{photo}.
%
%
The Curie temperature about 820K which was
observed in one of the samples \cite{narozhn} does not coincide
with those for cementite and $\alpha$-iron.
It is important to stress that, actually, only
magnetic carbon systems synthesized from ethanol
\cite{izr} can be considered as pure enough in a sense of
iron impurities whereas natural materials such as coal and graphite
always contain noticeable amount of iron and cannot be
completely purified \cite{Fegr1,Fegr2,Fegr3}. Thus, nanotubes
prepared from them are also contaminated by iron \cite{fecnt}
and iron oxides \cite{fe3o4} which are rather
difficult to eliminate \cite{analyst}. It is not surprizing
therefore that sometimes an amount of iron in fullerene
samples under discussion is as high as 400$\mu$gr/gr \cite{mak1,advmat}
but sometimes it is negligible \cite{condit}.
However, for the clearest samples in Ref.\cite{condit}
ferromagnetism was not observed so, indeed, it is a temptation
to connect the ferromagnetism of polymerized fullerenes
simply with some iron-based magnetic phases, such as elemental iron,
magnetite Fe$_3$O$_4$, or cementite Fe$_3$C \cite{Fe3C}.
However, this cannot explain the whole set of experimental data
available. First, magnetic characteristics of all these
phases are well known and usually do not coincide with those of
the ``magnetic carbon'' \cite{maknew}, besides that,
X-ray emission spectra of carbon in cementite are rather
peculiar which allows us always to distinguish cementite
from other carbon compounds \cite{Kurmaev}.
The magnetic measurements was used already
to separate intrinsic carbon magnetism from that of
``parasytic'' phases \cite{1} (for polymerized fullerenes it was done
in Ref. \cite{maknew}). Also, even intentional
introduction of iron to carbon systems is not
sufficient to produce the ferromagnetism \cite{Febomb,Fealone}.

In last few years new
results appear about ferromagnetism in heavy-ion irradiated
fullerene films \cite{irrad1,irrad2} and proton irradiated
graphite \cite{magnogr}. The x-ray magnetic circular dichroism
measurements demonstrate magnetic moments on carbon in proton
irradiated graphite \cite{XMCD1}.
Another argument for intinsic character of carbon
magnetism were discussed in Ref. \cite{C60H24} based on
experimental data on C$_{60}$H$_{24}$.
This provides some additional arguments in favor of
possibility of intrinsic carbon ferromagnetism under certain
conditions.
To summarize the experimental data discussed above,
one can conclude that there is a correlation between
contamination by iron and ferromagnetism in the polymerized
fullerenes \cite{rev} but it cannot be probably explained
by the most straightforward way, as a result of formation
of iron-based ferromagnetic phases. At least, some
experimental data seem to be in disagreement with this
view. Actually, the role of iron impurities in the problem
of magnetic carbon may be more subtle and some more
complicated opportunities can be considered, e.g.,
a formation of different molecular magnets
based on charged fullerenes, with the Curie temperature T$_C$
above room temperature (see Ref.\cite{Konarev} and references
therein). In prinicple, this might explain a variety of T$_C$
values in magnetic fullerenes as well as a fragility of
their ferromagnetism. However, these systems \cite{Konarev}
are periodic crystals and it is not known whether they
can be ferromagnetic, with high enough T$_C$, under the
disorder, or not. Also, they have specific Raman spectra
different from those observed in the magnetic fullerenes.

Another possible explanation is a catalytic activity of
transition metals, in particular, iron, and their oxides, which
are usually exploited for the nanotube gropwth
\cite{grow1,grow2,Oleg}. One can assume that iron impurities can
initiate a formation of various defects responcible for
the ferromagnetism. It is the scenario which will be discussed further.

Thus, the nature of magnetism in carbon based systems is still an
open issue and thus theoretical investigations can be very useful.
Based on first-principle calculations of electron energy spectra
and comparison with relevant experimental data it was shown
\cite{ourC60} that the ferromagnetism cannot be an intrinsic
property of ideal rhombohedral crystal lattice of the polymerized
fullerenes. A role of doping by light elements in the appearance
of ferromagnetism has been studied theoretically in Ref.
\cite{C60R}. A possible relation between vacancies and magnetism
in carbon based systems was investigated intensively in last
years. It is known \cite{Leht,Duplock,Yazyev1} that a single
vacancy in graphene produces a local magnetic moment about 1.2
$\mu_B$. At the same time, the vacancies in fullerenes were
studied rather in a context of their role in structure stability
of carbon cages \cite{vacancalc1,vacancalc2} and information
about their magnetic properties is rather poor. Magnetic
interactions between vacancies in the polymerized fullerenes were
studied in Ref. \cite{vacancalc3} but only for a one special kind
of vacancies. The model proposed in Ref. \cite{vacancalc3} cannot
explain the ferromagnetism of proton-irradiated samples
\cite{irrad1,irrad2}, as well as conservation of ferromagnetism
after the depolymerization \cite{photo}. At the same time, keeping
in mind strong suspicions about an extrinsic character of
``carbon'' magnetism a theoretical clarification of issues on
iron-carbon phases and metal impurities is necessary which was not
done yet.

Here we present the results of first principles calculations of
magnetic and structural properties of various types of vacancies
in fullerenes. It turns out that the recombination of magnetic
single vacancies into non-magnetic bivacancies (which have the
lower energy than a pair of the single vacancies) is the main
limiting factor. This probably explains appearance of the
ferromagnetism only in a narrow interval of pressure and
temperature \cite{condit} which was not discusses in the earlier
theoretical considerations. Based on our computational results we
present also some arguments against decisive role of iron
confirming that the ferromagnetism can be an intrinsic property of
(defected) carbon systems.

\section{Computational method}
\label{sec:1}

We used the SIESTA package for electronic structure calculations
\cite{siesta1,siesta2} with the generalized gradient approximation
for the density functional \cite{perdew} with energy cutoff
400 Ry, and $k$-point 8$\times$8$\times$8 mesh
in Monkhorst-Park scheme \cite{MP}. Coordinates of atoms and
lattice parameters were taken for perfect (defect-free) cubic
C$_{60}$ structure from Ref. \cite{coord1}, and for perfect
tetragonal and rhombohedral phases from Ref. \cite{coord2}.
Calculated electronic structure for both cubic and rhombohedral
C$_{60}$ are in a good agreement with the experimental X-Ray
spectra \cite{ourC60}.

\section{Iron impurities on C$_{60}$}
\label{sec:2}

\begin{figure*}
\resizebox{0.50\columnwidth}{!}{%
  \includegraphics{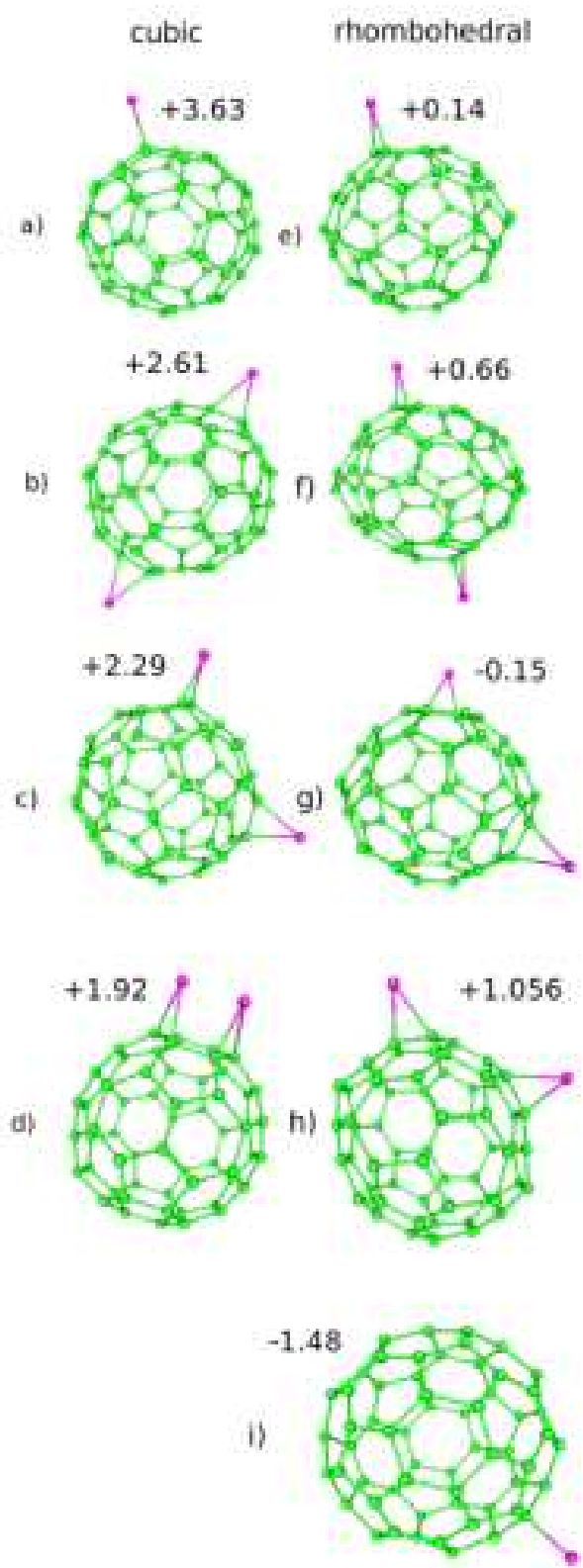}
}
\caption{Optimized geometrical
structure of iron atoms adsorbed on C$_{60}$ in cubic and
rhombohedral phases. Numbers near the pictures are defect
formation energies per iron atom, in eV.}
\label{fig1}       
\end{figure*}

First, we investigated possible segregation of Fe impurities. To
this aim, we have calculated solution (defect formation) energy
for two Fe atoms situated in the same C$_{60}$ ball and its
dependence on the distance between impurity for both cubic and
rhombohedral phase, with the optimization of atomic positions in
the buckyball. The lattice constants were kept fixed as it should
be in the case of small impurity concentration. The defect
formation energy was calculated as $E_{form} =
E_{C_{60}+nFe}-E_{C_{60}}+nE_{Fe}$, where $E_{C_{60}+nFe}$
is the energy of the ball C$_{60}$ with adsorbed $n$ atoms of iron,
$E_{C_{60}}$ is the total energy of pure C$_{60}$, and $E_{Fe}$
is the total energy of $\alpha$-Fe per atom.
Different geometric configurations
for these calculations and the calculated energies are shown in
Fig. \ref{fig1}. It is seen that for the cubic phase the single Fe
atom per C$_{60}$ ball has the highest energy whereas for the pair of
atoms the energy decreases with the distance decrease. This means
that iron impurities in the cubic phase will segregate, indeed.
However, for the rhombohedral phase the tendency is opposite. The
formation energy in this case is the lowest for the single atom
at the apex of the buckyball and for the pairs of Fe atoms it {\it
decreases} with the distance increase. For the atomic
configuration shown in Fig. \ref{fig1}g the formation energy is
smaller than for the single atom which is probably related with
geometric distortions of the C$_{60}$ sphere, the iron atom being
located in the void between neighboring buckyballs. We have
calculated additionally the formation energy for the single Fe
atom situated near the central part of rhombohedral carbon cage
(Fig. \ref{fig1}i). This corresponds to the minimal energy.
Thus, for the rhombohedral C$_{60}$ the segregation of Fe looks
energetically unfavorable. Instead, at high concentration of
impurities one can expect a formation of linear chains
metal-buckyball-metal-... which were experimentally observed for
the mixture of cobalt with C$_{60}$ \cite{Cochains}.

\section{Vacancies in C$_{60}$}
\label{sec:3}

\begin{figure}
\resizebox{0.75\columnwidth}{!}{%
  \includegraphics{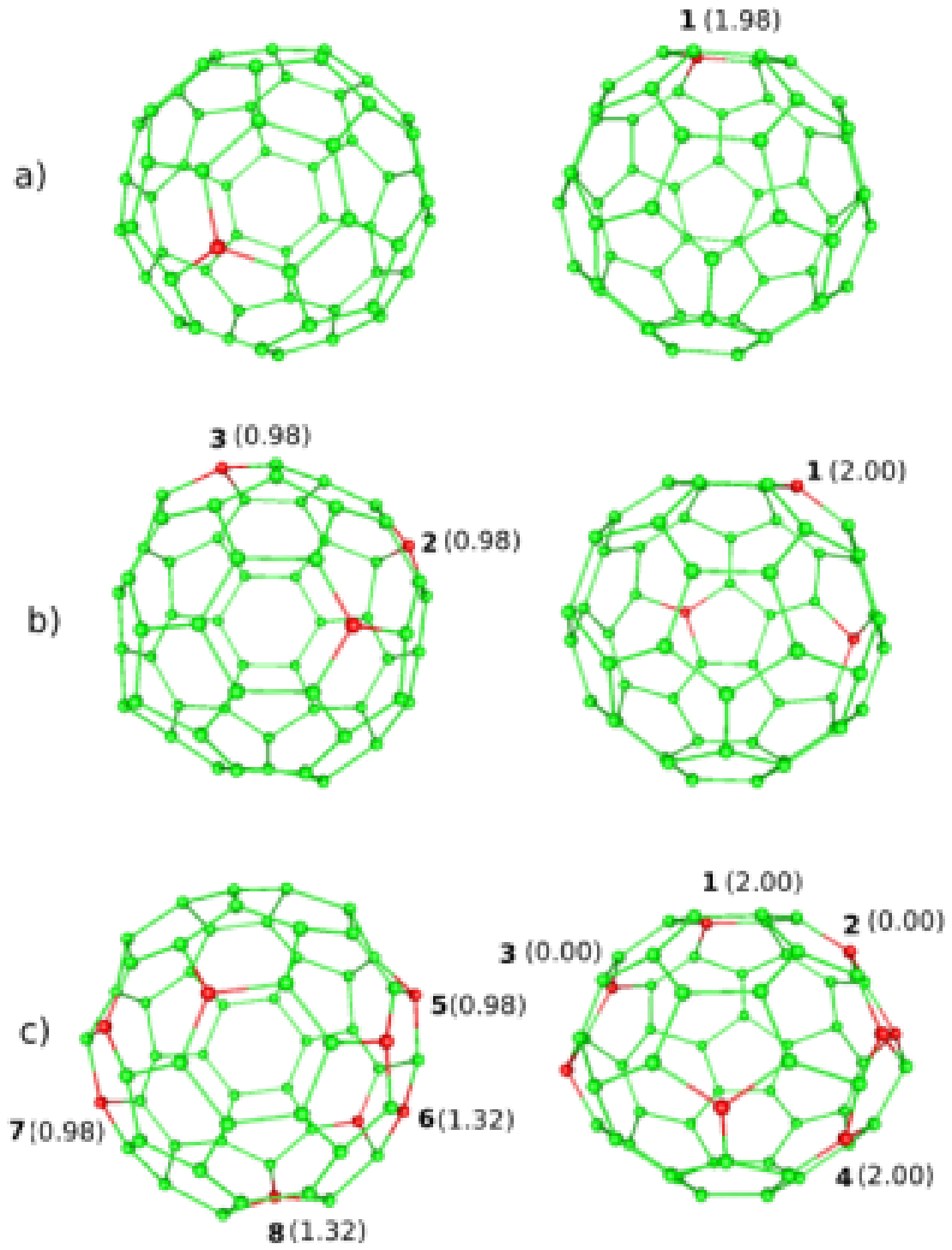}
}
\caption{Top (left) and side (right)
view of cubic (a), tetragonal (b) and rhombohedral (c) fullerenes.
Red (dark) spheres show inequivalent atoms which can be removed.
The values of magnetic moments per whole buckyball are shown in
parentheses (in $\mu_B$).}
\label{fig2}       
\end{figure}

The vacancy formation energy was calculated as $E_{form} =
E_{vac}-E_{C_{60}}(60-n)/60$, where $E_{vac}$ is the total energy of
the system with single vacancy per buckyball. As was shown in Ref.
\cite{vacancalc3} the vacancies can be responsible for
ferromagnetism of fullerenes. The main problem is their too high
energy formation, at least, in the cubic phase \cite{vacancalc1}.
However, the vacancy energy is sensitive to local geometric
structure of fullerenes and, as we will show, there is an
essential difference between cubic and rhombohedral phases. Figure
\ref{fig2} shows a pictorial view of the cubic, tetragonal  and
rhombohedral phases. Whereas all atoms in the cubic phase are
equivalent, already in the tetragonal one there are several
different types of vacancy positions, due to formation of covalent
bonds between the buckyballs and their distortions.  We have
performed calculations of the vacancy formation energy removing
nonequivalent single atoms as shown in Figure \ref{fig2}. The
calculated formation energy for the cubic phase is 5.58 eV which
is smaller than the value 7.2 eV for graphene \cite{vacan}.
In the tetragonal phase there are three
types of the vacancy positions, the formation energy being
decreased from the apex of the buckyball to its center. This is a
consequence of the carbon-carbon bonds between the buckyballs in
the central area which leads also to partial suppression of
magnetic moments of these vacancies.
The calculated formation energies for the
tetragonal phase vary from 4.67 eV to 6.28 eV.
Similar to Ref.\cite{vacancalc3} magnetic moments on vacancies are
strongly dependent on geometric structure of fullerene.

The rhombohedral C$_{60}$ is characterized by a larger number of
the C-C bonds and, as a result of polymerization, the ball is
compressed by 8.4\% along $Z$ axis and expanded in the
perpendicular directions and, thus, essentially different parts
can be separated, the central one and two ``capping'' parts. The
formation energy for vacancies in the capping parts (atoms 1 - 4
in Figure \ref{fig2}c) is about 5 eV which is close to
corresponding values for other C$_{60}$ phases. However, for the central
part (atoms 5 - 8 Figure \ref{fig2}c) the vacancy
formation energy is much smaller and for the atoms 5 and 7 it
reaches 3.22 eV. That is why, probably, the polymerization at high
pressure and high temperature is the way to magnetic carbon: it
diminishes the vacancy formation energy and thus leads to a drastic
increase in the vacancy concentration.

The magnetism on the vacancies turns out to be rather fragile. The
broken bonds required for magnetism are, at the same time,
active centers for chemisorption. As a result, the magnetic
moments can be destroyed by impurities. We have calculated the
energy of oxygen chemisorption on a single vacancy. It is equal to
-1.114 eV  for the most energetically favorable vacancy in the
rhombohedral case. At the same time, for the vacancy in cubic
phase this energy is {\it positive} and equals 0.716 eV. These
data can explain suppression of ferromagnetism at the
photopolimerization of C$_{60}$ under oxygen-rich conditions
\cite{photo}.

\begin{figure}
\resizebox{0.75\columnwidth}{!}{%
  \includegraphics{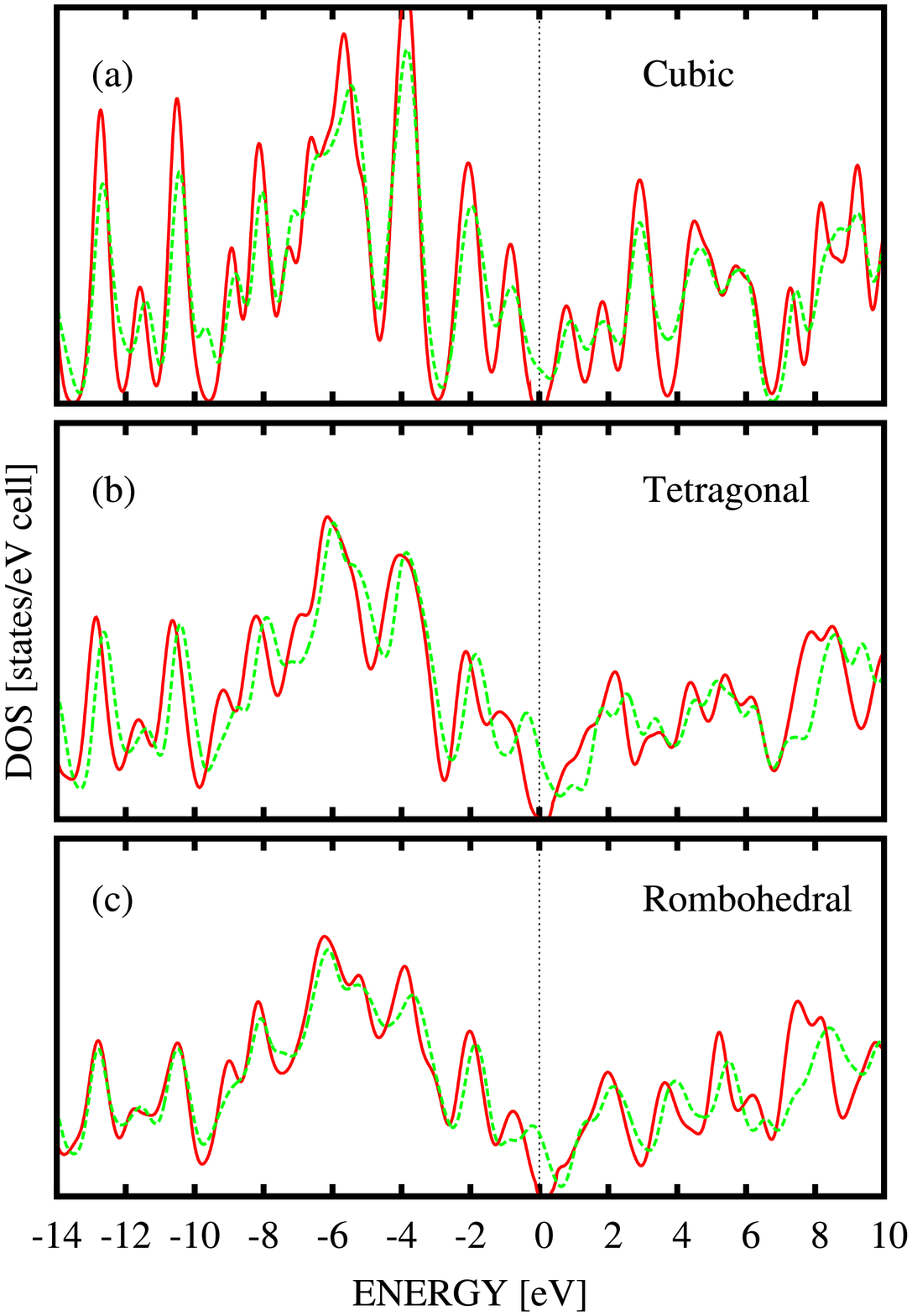}
}
\caption{Total densities of states for
(a) cubic, (b) tetragonal, (c) rombohedral fullerens. Solid red
and dashed green lines correspond to the system without and
with vacancy, respectively.}
\label{fig3}       
\end{figure}

Figure \ref{fig3} presents electron densities of states for all
studied phases of C$_{60}$ with and without vacancies. At a large
energy scale, vacancies modify the electronic structure rather
weakly so it would be difficult to see their effects in, say,
X-ray spectra \cite{ourC60}. At the same time, they change
essentially the electronic structure in the close vicinity of the
Fermi level producing conduction electrons in the former energy
gap. Note that experimentally magnetic samples turned out to be
conducting \cite{conduct}. These charge carriers may be
responsible for ferromagnetism appearance in a narrow defect band
\cite{EK}.

\begin{figure}
\resizebox{0.75\columnwidth}{!}{%
  \includegraphics{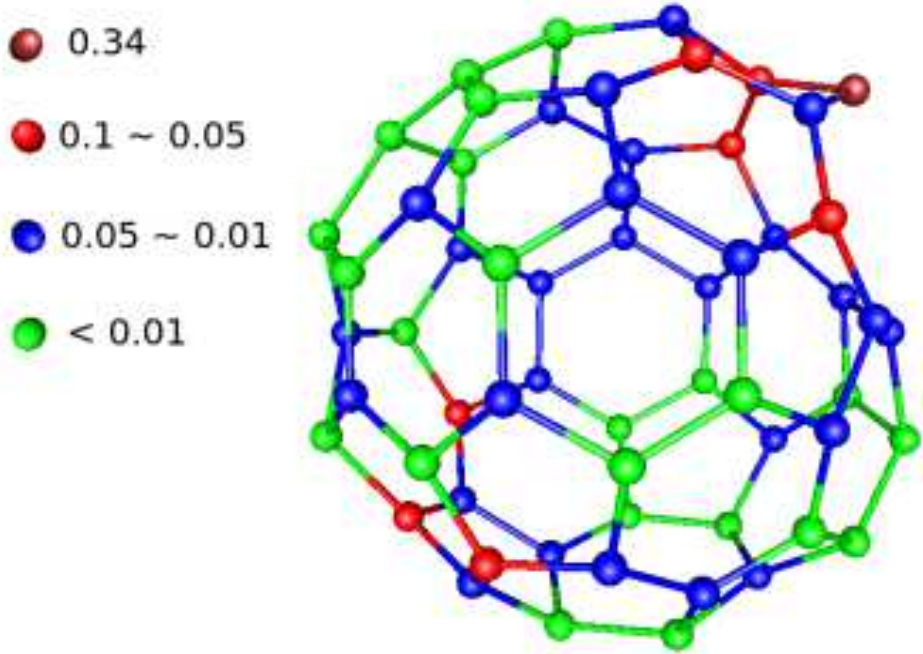}
}
\caption{Distribution of magnetic moments in $\mu_B$
on C$_{60}$ sphere with single vacancy in rhombohedral phase.}
\label{fig4}       
\end{figure}

To estimate the exchange interaction energy responsible for
possible magnetic ordering we have performed a total energy
calculation for a supercell containing two C$_{60}$ balls with
single vacancy per ball, with parallel (FM) and antiparallel (AFM)
spin orientations. The calculations have been done for both cubic
and rhombohedral phase, in the latest case the vacancies with the
lowest formation energy being considered. For both systems the
atomic relaxation has been taken into account. We have found that
the difference between total energy in FM and AFM configurations
is 0.41 eV for the rhombohedral phase and 0.53 eV for the cubic
one. These values are close to those calculated for the
hydrogen-doped C$_{60}$ \cite{C60R}. This energy difference is
very high which can be explained by a peculiar distribution of the
electron spin density. Whereas in open structures like graphene
the magnetic moments from vacancies and impurities are rather
localized, with a fast decay of the spin density \cite{Pisani,Yazyev1},
in closed structures such as C$_{60}$ the magnetic
moments distribute over the whole buckyball (see Figure
\ref{fig4}, cf. Ref. \cite{C60R} for the hydrogen-doped
fullerenes). Of course, the value of exchange interactions between
distant balls with the vacancies will be weaker than that between
the neighboring ones so our calculation gives just an upper limit
for the value of magnetic interactions.
Various concentration of fullerenes with vacancies will
result in various distances between them and, thus, with
various exchange parameters and Curie temperatures which
seems to be in agreement with experimental data showing
a large variance of T$_C$ in polymerized fullerenes,
from 550 K to 800 K.

\begin{figure}
\resizebox{0.75\columnwidth}{!}{%
  \includegraphics{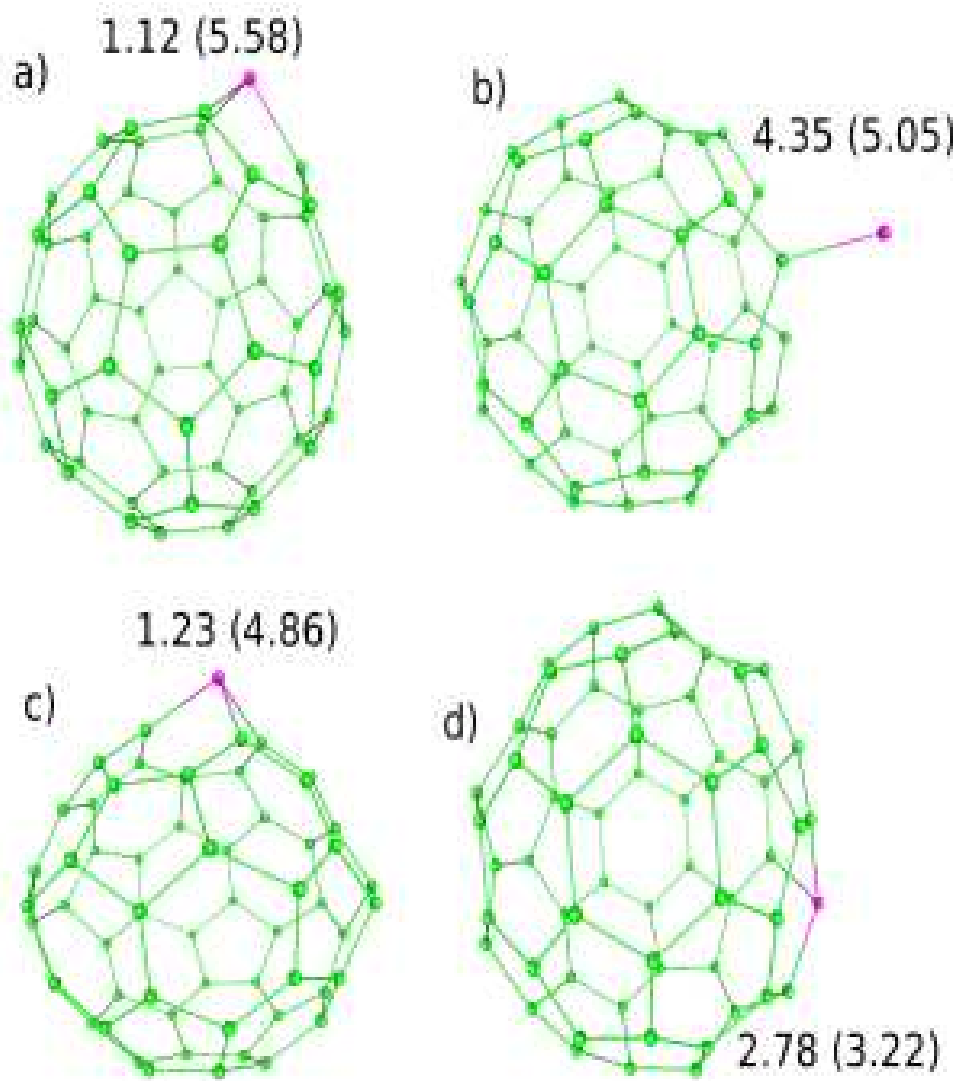}
}
\caption{Optimized structures of fullerenes with iron impurity
and carbon vacancy nearby for cubic phase (a) and nonequivalent
positions in rhombohedral (b)-(d). The numbers correspond
to formation energies of the vacancy in electronvolts with
(without) Fe impurity.}
\label{fevac}       
\end{figure}

Let us consider now an effect of iron impurities on
the vacancy formation in fullerenes. By analogy with the process
of ripping of few layer graphene by iron
nanoparticles \cite{cut3} the formation
energy will be calculated by the expression
$E_{form} = E_{C_{60}Fe} - E_{C_{59}Fe} - E_{C_{60}}59/60$.
As the simplest model we consider the system $C_{60}Fe$
where carbon atom closest to the iron impurity is removed.
We start with the cubic phase of fullerene.
The computational results are presented in
Fig.\ref{fevac}a. One can see that in the presence of Fe
the vacancy formation energy is drastically decreased.
Since the iron atoms in cubic phase tend to segregate as
was discussed above this caluclation deals with rather
hypopthetic situation than real one. However, a similar tendency
holds also for the polymerized fullerenes.
As shown in Fig.\ref{fevac}b-d for all positions
of Fe under consideration the vacancy formation
energy is essentially lower than without iron.
Because single iron atoms without magnetic
carbon impurities in neighbors stay paramagnetic centers
\cite{Fealone}.
Thus, contamination by iron can initiate a formation of
defects responsible for ferromagnetism.
The effect of Fe on the vacancy formation
cosidered here may be also interesting in light of
simulation of metal-fullerene systems for potential
application in hydrogen storage \cite{Sc,Ti}.

\begin{figure}
\resizebox{0.75\columnwidth}{!}{%
  \includegraphics{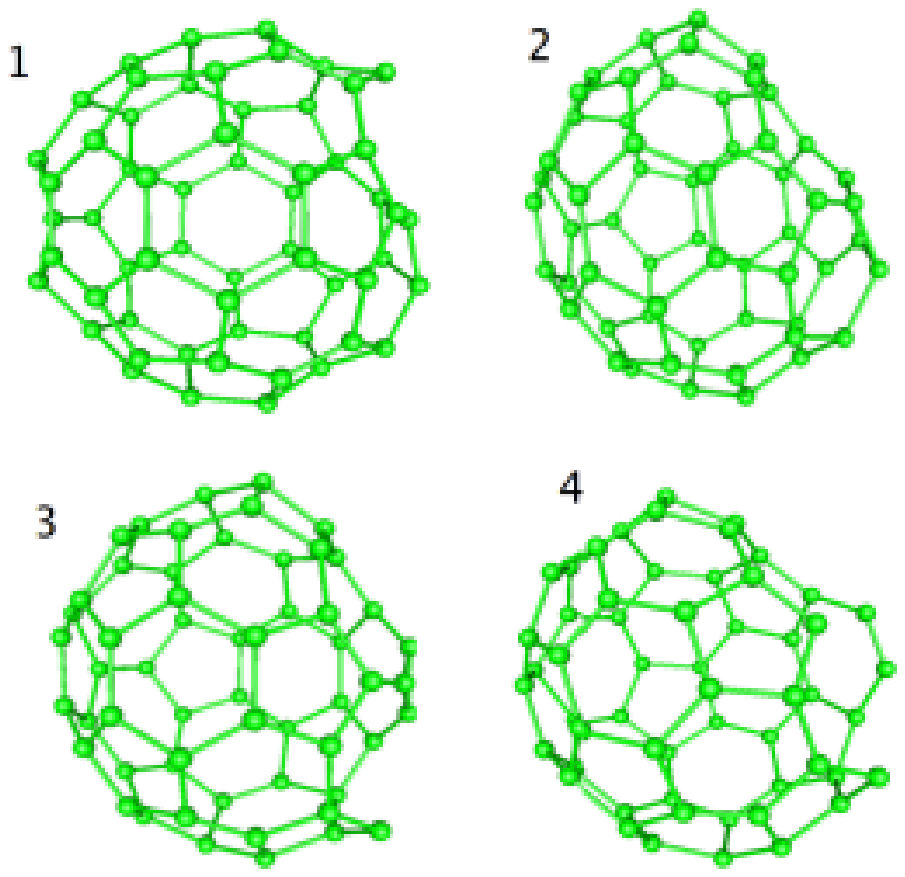}
}
\caption{Optimized geometrical structures of
 mono-, bi-, tri- and quadrovacancies in rhombohedral
C$_{60}$.}
\label{fig5}       
\end{figure}

\begin{figure}
\resizebox{0.75\columnwidth}{!}{%
\rotatebox{-90}{
  \includegraphics{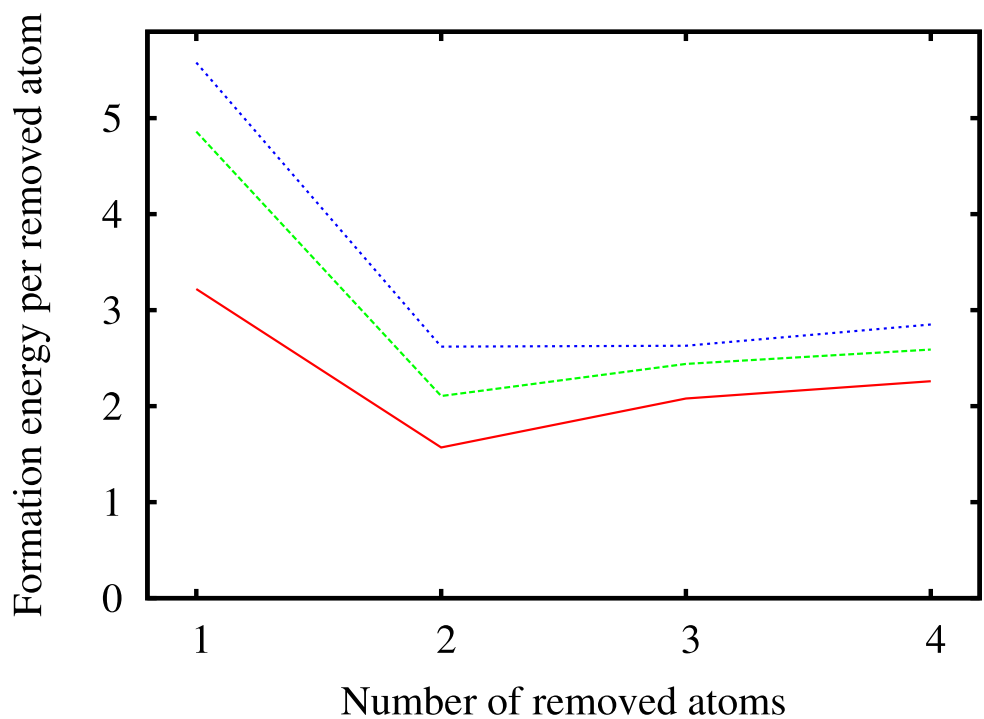}
}}
\caption{Energy formation for
multivacancies per removed atom as a function of the number of
removed atoms in central part of rhombohedral C$_{60}$ (solid
red line), on top of the rhombohedral C$_{60}$ (dashed green
line), and in cubic C$_{60}$ (dotted blue line).}
\label{fig6}       
\end{figure}

Based on the computational results one can conclude that vacancies
can be a cause for the ferromagnetism of polymerized fullerenes.
To understand, in a framework of this picture, why the
ferromagnetism is so fragile and why it disappears at the
temperature of pressure increase the energetics of multivacancy
formation has been considered. We have calculated the energies and
magnetic moments for bi-, tri- and quadrovacancies for both cubic
and rhombohedral C$_{60}$ (see Fig. \ref{fig5}).
The computational results presented in
Fig. \ref{fig6} demonstrate that the formation energies per
removed atom for the bivacancies are smaller than for the single
vacancy, similar to the previous results for graphene
\cite{vacan}. Therefore, the process of vacancy fusion into the
bivacancy is energetically profitable. The bivacancy is always
nonmagnetic since there is no unpaired electrons in the situation
of {\it two} broken bonds. This explains a very narrow temperature
and pressure interval where the ferromagnetism in polymerised
C$_{60}$ arises: further increase of the pressure or temperature
stimulates the process of the bivacancy formation. Figure
\ref{fig6} shows that the appearance of the multivacancies is the
most probable in the central part of rhombohedral C$_{60}$. This
is exactly the place where the cage collapse happens as is
confirmed by observation of curved fragments of fullerenes in
the ``hard carbon'' after the cage collapse \cite{hard1,hard2}.

\section{Conclusion}
\label{sec:4}

Thus, the explanation of appearance and disappearance
of ferromagnetism in polymerized fullerenes
in terms of vacancies seems to be consistent with the whole set of
experimental data. According to this picture, the ferromagnetism
arises under rather restrictive conditions where the monovacancies
are formed but their fusion into nonmagnetic bivacancies can be
prevented. Oxygen chemisorption will also suppress the
ferromagnetism due to formation of chemical bonds involving
unpaired electrons associated with the vacancies. We have
discussed also mechanisms of the cage collapse via fromation of
quadrocvacancy in the central part of buckyballs. We have
demonstrated that a segregation of iron impurities in the
rhombohedral fullerene is energetically unprofitable, in contrast
with the cubic one.
Importantly, a contamination by iron can decrease the vacancy
formation energy in fullerenes.

The work is financially supported by Stichting voor
Fundamenteel Onderzoek der Materie (FOM), the Netherlands.

\end{document}